# An Explainable Neural Radiomic Sequence Model with Spatiotemporal Continuity for Quantifying 4DCT-based Pulmonary Ventilation

Rihui Zhang[1,2], Haiming Zhu[1,2,3], Jingtong Zhao[4], Lei Zhang[1,2], Chunhao Wang[4], Fang-Fang Yin[1,2], and Zhenyu Yang[1.2,*]

1. Medical Physics Graduate Program, Duke Kunshan University, Kunshan, Jiangsu, China
2. Jiangsu Provincial University Key (Construction) Laboratory for Smart Diagnosis and Treatment of Lung Cancer, Kunshan, Jiangsu, China
3. Department of Radiotherapy and Oncology, The First People's Hospital of Kunshan, Kunshan, Jiangsu, China
4. Department of Radiation Oncology, Duke University, Durham, NC, United States
5. Department of Radiology, The First People's Hospital of Kunshan, Kunshan, Jiangsu, China

**Short Running Title**: *Neural Radiomic Sequence for 4DCT Ventilation*

*Corresponding author:

Zhenyu Yang, Ph.D.
Medical Physics Graduate Program
Duke Kunshan University
Jiangsu, China

E-mail: zy84@duke.edu

**Abstract**

**Background:** Regional lung ventilation assessment is critical for managing and treating patients with lung diseases. While nuclear medicine-based ventilation scintigraphy remains the clinical standard, it is time-consuming, costly, and involves additional radiation exposure.

**Purpose:** To develop an explainable neural radiomic sequence model for voxel-wise identification of impaired pulmonary ventilation using 4DCT imaging.

**Methods:** A total of 45 lung cancer patients from the VAMPIRE dataset were analyzed, where 25 had PET/4DCT, and 20 had SPECT/4DCT acquired. For each subject, lung volumes were segmented across respiratory phases of 4DCT, and 56-dimensional voxel-wise radiomic features were extracted throughout the respiratory cycle to construct temporal radiomic sequences encoding local intensity and texture dynamics. A temporal saliency-enhanced long short-term memory (LSTM) network was developed to model spatiotemporal continuity while providing explainable predictions. Ground-truth ventilation defect labels were derived from Galligas-PET or DTPA-SPECT by thresholding method. The model was trained and evaluated via five-fold cross-validation, with performance assessed using the Dice similarity coefficient (DSC), sensitivity, accuracy, and area under the ROC curve (AUC-ROC). Temporal saliency maps were generated to identify key contributing radiomic features and respiratory phases. Comparative experiments were conducted against two baseline models: U-Net and LSTM trained directly on 4DCT images.

**Results:** The proposed model outperformed both baselines, achieving mean (range) DSC scores of 0.78 (0.74–0.79) in PET cohort and 0.78 (0.74–0.82) in SPECT cohort, compared to 0.51 (0.44–0.55)/0.51 (0.40–0.58) for U-Net model and 0.69 (0.65–0.74)/0.66 (0.60–0.70) for LSTM model for the two cohorts respectively. Similarly, mean AUC-ROC values were 0.85 and 0.84 for PET and SPECT cohorts, respectively, using our model, versus 0.68/0.65 using U-Net) and 0.77/0.71 using LSTM. Accuracy and sensitivity showed consistent improvements. Saliency analysis highlighted three radiomic sequence patterns characterizing ventilation dysfunction: impaired regions exhibited (1) increasing intensity and (2) decreasing homogeneity during exhalation—dynamics that contrast sharply from healthy lung tissue.

**Conclusion:** This work demonstrates the feasibility of radiomic sequence modeling for functional lung assessment using 4DCT and potentially offers a non-invasive, explainable alternative to conventional ventilation imaging. By explicitly modeling radiomic feature evolution across the respiratory cycle, our proposed framework preserves pulmonary motion dynamics and enables voxel-wise quantitative characterization of ventilation.

## 1. Introduction

Pulmonary diseases, including chronic obstructive pulmonary disease (COPD), asthma, pulmonary fibrosis, and lung cancer, significantly impact global health, ranking as major contributors to morbidity and mortality worldwide [1-4]. Accurate assessment of lung function is crucial in the clinical management of these conditions [5, 6]. Ventilation, the exchange of air in and out of the lungs, is the most common surrogate of lung function [7-9]. The identification of ventilation defects plays a key role in disease management [10], functional avoidance radiotherapy treatment planning [11, 12] and regional therapeutic response evaluation [4, 13]. Traditional pulmonary ventilation assessments, including forced expiratory volume in one second ($FEV_1$) [14] and diffusing capacity of the lung for carbon monoxide (DLCO) [15], measure the total volume and flow of air inhaled and exhaled, providing a global quantification of lung ventilation [16]. These assessments are rapid, easy and commonly utilized in routine ventilation evaluations [17]. However, these techniques lack the ability to locate ventilation abnormalities, limiting their utility in spatially resolved clinical interventions [16]. In response, imaging-based techniques for regional ventilation quantification have gained increasing interest.

Positron Emission Tomography (PET) using Galligas—an inhaled Ga-68-labeled radioactive aerosol—enables voxel-level visualization of air distribution in the lungs [18, 19]. Similarly, Single Photon Emission Computed Tomography (SPECT) employs inhaled Technetium-99m-labeled Diethylenetriamine Pentaacetic Acid (DTPA) to generate regional ventilation maps [19-21]. Hyperpolarized noble gas (HNG) magnetic resonance imaging (MRI) offers another avenue [22], using hyperpolarized helium-3 [23] or xenon-129 [24] to produce high-resolution images of lung ventilation [25, 26]. Xenon-enhanced CT has also shown promise in feasibility studies for delineating ventilation patterns based on contrast-enhanced gas distribution [27]. While effective, these methods require specialized tracers and equipment, involve lengthy acquisition protocols, and may expose patients to additional radiation and high costs [12, 28].

To address these limitations, Computed Tomography Ventilation Imaging (CTVI) techniques have been developed to estimate ventilation directly from four-dimensional CT (4DCT) scans [10, 12, 29]. Unlike static CT, 4DCT captures lung motion throughout the respiratory cycle [30], enabling dynamic modeling of lung deformation. CTVI methods rely on deformable image registration (DIR) to align lung volumes across respiratory phases [31-33], generating displacement vector fields (DVFs) that describe voxel-level motion [9, 10, 32]. Ventilation is then inferred using the Jacobian determinant of DVFs (DIR-JAC) or through Hounsfield Unit changes between registered voxels (DIR-HU), with larger local volume changes typically interpreted as higher ventilation activity [9, 31]. However, the reliability of CTVI remains dependent on DIR algorithm accuracy and the precision of lung motion modeling [10, 34-37]. Therefore, achieving robust and reproducible CTVI results remains challenging across heterogeneous populations and imaging protocols [9, 10, 37].

Recent advances in machine learning and computational power have introduced radiomics as a novel, DIR-independent approach to functional lung imaging [38]. Radiomics refers to the extraction of quantitative imaging features—including intensity, texture, and shape—based on predefined mathematical formulations [39, 40]. These handcrafted features may serve as biomarkers that reflect physiological and pathological properties of lung tissues [41, 42]. Machine learning models trained on radiomic features have shown promise in predicting global pulmonary function metrics such as $FEV_1$ and DLCO [43]. Our prior work extended this framework by employing voxel-wise radiomic filtering over 4DCT scans to characterize regional variations in lung function [44]. Preliminary findings indicated that radiomic feature maps derived from average 4DCT images may correlate with regional ventilation patterns [44]. Parallel efforts using deep neural networks (DNNs) have demonstrated the feasibility of predicting ventilation distributions directly from CT or 4DCT images via supervised learning [36, 45]. Despite their potential, existing radiomics and DNN-based approaches often rely on static CT imaging, neglecting the dynamic nature of respiratory motion and the temporal evolution of imaging features. Furthermore, the interpretability of such models remains limited. DNNs learn hierarchical feature representations through nonlinear transformations, resulting in models that are often opaque or "black-box" in nature [46, 47]. For example, the specific locoregional texture patterns and spatiotemporal lung motion contributing to the predicted ventilation distributions remain unknown, thereby impeding clinical trust and adoption.

In this study, we propose an explainable neural radiomic sequence model that incorporates spatiotemporal continuity to identify compromised pulmonary ventilation regions from 4DCT images. Specifically, locoregional intensity and texture features were extracted across the respiratory cycle, forming radiomic sequences that capture time-dependent variations in lung tissue characteristics. These sequences were analyzed using a long short-term memory (LSTM) network—an architecture well-suited for modeling temporal dependencies in sequential data [48, 49]. An explainability module was further embedded to derive temporal saliency maps, enabling identification of key radiomic sequences and respiratory phases contributing to the model's decision. Validation was performed against Galligas PET and DTPA-SPECT ventilation images. To our knowledge, this is the first work that integrates spatiotemporal radiomic dynamics into an explainable deep learning framework for ventilation quantification. By leveraging routinely acquired 4DCT data [33, 50], our approach offers a non-invasive, no additional radiation, and explainable alternative to existing ventilation imaging techniques.

The main contributions of our work can be summarized as follows:

- We extend traditional radiomics from static imaging to dynamic 4DCT, enabling time-resolved modeling of lung tissue heterogeneity through radiomic filtering across all respiratory phases. This is the first implementation of motion-aware radiomic sequence modeling for ventilation estimation.
- We introduce a novel explainable neural radiomic sequence model that accurately identifies regions of impaired ventilation, achieving high average Dice scores of 0.78/0.78 (PET/SPECT cohorts) and AUC-ROC values of 0.85/0.84.
- Our model generates intuitive temporal saliency maps that reveal the importance of specific radiomic features and respiratory phases in predicting ventilation defects. These maps provide clinically explainable insights into the spatiotemporal mechanisms driving model predictions.
- Given the broad clinical adoption of 4DCT, the proposed method offers a practical and generalizable solution for functional lung imaging without the need for additional imaging modalities or contrast agents.

## 2. Related work

This section conducts a literature review for CTVI, radiomics/deep learning-based ventilation imaging, time series modelling, and the explainability of time series models.

### *2.1. Computed tomography ventilation imaging*

Currently, CTVI primarily relies on two methodological frameworks: DIR-JAC and DIR-HU approaches. Both begin by registering the inhalation and exhalation phases of 4DCT using a DIR algorithm, resulting in a DVF that characterizes voxel-wise lung deformation across the respiratory cycle [10]. The DIR-JAC method estimates local volume changes by computing the Jacobian determinant of the DVF at each voxel. The Jacobian matrix, formed from the spatial gradients of the displacement field, quantifies local volumetric transformations through its determinant. A value greater than one indicates local expansion (i.e., increased air volume), while values less than one reflect local compression. These volumetric strain metrics are then interpreted as surrogates for regional ventilation, under the assumption that greater volume expansion corresponds to increased airflow and greater ventilation [9, 45]. In contrast, the DIR-HU method infers ventilation based on changes in CT intensity, i.e., HUs, between spatially aligned voxels following DIR [9, 51]. As inhalation increases lung air content and reduces tissue density, a decrease in HU is interpreted as local expansion. This HU difference is converted into a quantitative estimate of volume change, thereby serving as a proxy for ventilation. Notably, this method does not directly rely on the differential geometry of the DVF but instead assumes that tissue density changes sufficiently reflect ventilation dynamics in regions where registration accuracy is preserved.

Several comparative studies have evaluated the performance of these two approaches. Castillo et al. [52] reported similar accuracy between DIR-JAC and DIR-HU in estimating ventilation, suggesting that either method can be viable under appropriate conditions. Keall et al. [53] compared both methods against Technegas-based SPECT imaging and concluded that the fidelity of CTVI predictions is highly sensitive to 4DCT image quality, particularly motion artifacts and noise, which directly influence DIR accuracy and HU stability. The VAMPIRE Challenge is a large-scale benchmark study that tests 37 CTVI algorithms using a standardized 4DCT dataset [9]. The substantial variability in algorithmic performance has been observed, and this variability was largely attributed to differences in DIR algorithms and parameter settings across methods. A systematic review by Hegi-Johnson et al. [10] further highlighted that a fundamental limitation of current CTVI techniques lies in their strong dependence on the accuracy of DIR and the precision of lung motion modeling.

*2.2. Radiomics/Deep Learning-based ventilation imaging*

With recent advances in computational power and algorithmic development, quantitative techniques such as radiomics and deep learning have been increasingly applied to extract clinically meaningful patterns from medical images. In radiomics-based approaches, analysis typically begins with delineating a predefined volume-of-interest (VOI)—such as the lung parenchyma in ventilation studies—followed by the extraction of a set of handcrafted features defined by domain expertise. These features quantitatively characterize the VOI in terms of intensity distribution, shape and geometry, size or volume, and texture heterogeneity [54, 55]. The resulting feature vectors serve as potential imaging biomarkers that may capture subclinical manifestations of disease and correlate with functional metrics such as regional or global lung ventilation [38, 43, 44]. Several studies have demonstrated the utility of radiomics for pulmonary ventilation estimation. Lafata et al. [43] identified statistically significant associations between radiomic features extracted from whole-lung CT and global pulmonary function tests, including forced expiratory volume in one second (FEV1) and diffusing capacity of the lung for carbon monoxide (DLCO). Building on this concept, Yang et al. [44] extracted spatially encoded intensity and texture radiomic features from sub-regions of the averaged 4DCT lung and demonstrated their correlation with regional ventilation distributions. Similarly, Westcott et al. [38] applied a support vector machine classifier to predict regional ventilation using texture-based radiomic features extracted from 3D CT scans of COPD patients.

In parallel, deep learning has emerged as a powerful alternative for automated image analysis and feature learning [56]. Unlike handcrafted features, deep learning methods employ multi-layer neural networks that learn hierarchical representations directly from raw image data. When trained in a supervised manner using medical images paired with ground truth ventilation images—such as PET or SPECT—deep neural networks can model complex, nonlinear mappings from structural to functional imaging domains [57, 58]. Among these, convolutional neural networks (CNNs) have become the popular architecture in ventilation quantification [55, 59]. Zhong et al. [57] and Liu et al. [45] developed CNN models for generating ventilation images directly from 4DCT, demonstrating improved agreement with ground-truth imaging compared to traditional DIR-HU and DIR-JAC methods. Kajikawa et al. [60] further extended this concept by training a U-Net to translate CT images into synthetic SPECT ventilation maps, while also incorporating model uncertainty quantification.

Despite these promising results, current radiomics and deep learning approaches have notable limitations. Most existing methods do not fully leverage the continuous nature of respiratory motion captured by 4DCT, instead relying on static or averaged representations that may underrepresent temporal dynamics. Interpretability is another major concern in clinical applications. The internal operations in modern deep learning models involve complex nonlinear transformations across multiple layers, making it difficult to trace specific input features or temporal cues that drive output predictions [61].

### *2.3. Times series modelling*

Time series data refers to a sequence of observations indexed in temporal order, where each data point corresponds to a specific time step [62-65]. Due to its inherently high dimensionality and temporal continuity, time series data is typically modeled as an integrated structure rather than a collection of isolated values [63]. Analytical tasks on time series primarily focus on capturing temporal dependencies—both short-range and long-range correlations—embedded within the sequential structure [66].

Various deep learning architectures have been specifically designed for temporal modeling, including recurrent neural networks (RNNs), long short-term memory networks (LSTMs), bi-directional LSTMs (BiLSTMs), and Transformers. The RNNs model includes cyclic connections, which enable these networks to update their current state by integrating information from past states and current input data. Such recurrence mechanism has been widely proven to facilitate the modeling of temporal dependencies in sequential data [49, 67, 68]. The standard RNN model can be limited by vanishing and exploding gradient problems encountered during backpropagation through time, which hinder their ability to capture long-term dependencies [49]. To address these issues, the long short-term memory (LSTM) network was introduced as an enhanced RNN architecture [48]. LSTMs incorporate memory cells and gated control mechanisms—specifically input, output, and forget gates—that regulate the information flow within the network. These gating mechanisms enable the selective retention or discarding of information, making LSTMs effective in capturing short- and intermediate-term temporal dependencies [69]. The bi-directional LSTM (BiLSTM) architectures extend the standard LSTM by processing sequences in both forward and backward directions, allowing the model to exploit both past and future context simultaneously [70].

More recently, Transformer architectures have emerged as a powerful alternative for sequence modeling. Originally developed for natural language processing, the Transformer employs a non-recurrent encoder–decoder architecture based on multi-head self-attention mechanisms and position-wise feed-forward layers [71]. The encoder maps the input sequence into a latent representation, while the decoder generates the output sequence by selectively attending to relevant components of the encoded input. Unlike RNN-based models, the Transformer processes entire sequences in parallel, enabling highly efficient computation and the capacity to model global dependencies without recurrence [71]. The self-attention mechanism inherently captures long-range interactions by assigning context-dependent weights across all time steps, which has proven particularly advantageous for long-sequence modeling. However, the architectural complexity of Transformers, combined with their large parameter space, imposes significant challenges in terms of convergence stability and data efficiency, often requiring extensive training data for optimal performance.

### *2.4. Explainability in time series models*

Explaining deep learning models for time series data remains a significant challenge due to the intertwined nature of temporal dependencies and multivariate feature interactions. Efforts have been made to extend the Gradient-based saliency methods—such as Vanilla Saliency [72], Integrated Gradients (IG) [73], and SmoothGrad [74]—into time series models. These methods generate visual or numerical attributions that aim to quantify the contribution of each input component to the model's prediction. Vanilla Saliency computes the gradient of the output class score with respect to the input image to evaluate the sensitivity of the model's prediction to pixel-level changes [75, 76]. The resulting heatmaps, referred to as saliency maps, visually represent the importance of each pixel in the model's decision-making process. IG calculates the importance of each pixel by accumulating gradients along a straight-line path from a reference input to the target input [73]. SmoothGrad is a refined variant of Vanilla Saliency and calculate saliency maps by averaging gradients obtained from multiple noisy perturbations of the input [74]. When applied directly to time series models, however, these methods often struggle to account for the temporal structure of the data. Specifically, standard saliency approaches may produce misleading attributions by failing to disentangle temporal causality from feature relevance [77, 78].

In response, the Temporal Saliency Rescaling (TSR) framework has been proposed to adapt IG for time series data [77]. TSR decomposes the saliency attribution process into two components: (1) *Time Relevance Score*: For each time step, the saliency contribution is quantified by measuring the change in total attribution when that specific time step is masked or occluded; and (2) *Feature Relevance Score*: For each feature within a given time step, the saliency contribution is measured by assessing the change in attribution when that feature is masked. The final joint importance score for each (time, feature) pair is computed as the product of the time relevance and feature relevance scores. This factorized approach enables more accurate identification of both temporally and semantically salient patterns in multivariate time series, yielding explainable insights into the decision-making behavior of temporal deep learning models [77]. By leveraging TSR, IG-based saliency methods can be adapted to provide fine-grained, high-fidelity explanations in time series applications.

# 3. Methods

## 3.1. Radiomic sequence modelling

The overall design of the proposed radiomic sequence modelling for the lung ventilation quantification is shown in Fig. 1. The 3D lung volume at each phase of the 4DCT image was first segmented from the lung CT image, as shown in Fig. 1(A). The radiomic filtering technique was systematically applied across all respiratory phases using a 3D sliding window approach to capture locoregional intensity and texture patterns throughout the entire lung volume, as shown in Fig. 1(B). Specifically, for each 4DCT phase, a predefined 3D kernel traversed the lung volume with single-voxel step precision. At each lung voxel tomographic coordinate, a cubic sub-volume was defined to extract radiomic intensity and texture features. Each voxel coordinates within the 3D lung volume thus can be represented as an $n$-dimensional feature vector, and the feature space can be represented as a set of 3D feature maps. As illustrated in Fig. 1(B), the radiomic map retains the same matrix dimensions as the corresponding CT images. Such radiomic filtering procedure was applied consistently across the entire respiratory phases. Therefore, for a given radiomic feature at a given lung voxel tomographic coordinate, the evolution of radiomic feature values throughout the respiratory cycle can be conceptualized as a spatiotemporal-continuous *radiomic sequence* $\Phi$. The dynamic changes in $\Phi$ represent the evolution of locoregional lung intensity and texture along respiratory motion and deformation. The red and pink waves in Fig. 1(B) and (C) showed two examples of the obtained radiomics sequences for a given voxel tomographic coordinate in the left upper lung.

Mathematically, let $\mathbf{T}$ represents the collection of all $T$ time step during respiratory cycle, i.e., $\mathbf{T} = 1, 2, \ldots, T$, and $\mathbf{N}$ be the collection of all $N$ extracted radiomic features, i.e., $\mathbf{N} = 1, 2, \ldots, N$. For a given lung voxel tomographic coordinate $(x, y, z)$, $f_n^t$ is the value of feature $n$ ($n \in \mathbf{N}$) at time $t$ ($t \in \mathbf{T}$). The radiomic feature vector at time $t$, denoted $\theta_t$, can be formally represented as:

$$\theta_t = \{f_n^t\}_{n \in \mathbf{N}} \tag{1}$$

The spatiotemporal-continuous radiomic sequence for feature $n$, denoted $\Phi_n$, can be defined as:

$$\Phi_n = \{f_n^t\}_{t \in \mathbf{T}} \tag{2}$$

Therefore, each voxel at $(x, y, z)$ can be characterized by $N$ radiomics sequences, i.e., $\mathbf{\Phi} = \{\Phi_n\}_{n \in \mathbf{N}}$. Therefore, the feature space $\boldsymbol{\mathcal{F}}$ for each patient can be represented as:

$$\boldsymbol{\mathcal{F}} = (\boldsymbol{\Phi}_{xyz}) \in \mathbb{R}^{\mathrm{X}\times\mathrm{Y}\times\mathrm{Z}} \tag{3}$$

where $X \times Y \times Z$ is the 3D space coordinate of the entire lung volume.

In this study, a total of $N = 56$ radiomic features were studied to capture the locoregional lung intensity and texture characteristics [79, 80]. These features can be summarized by three categories:

1) 18 intensity-based features: quantify pixel intensity distribution across the image.
2) 22 gray level co-occurrence matrix (GLCOM)-based features: describe fine texture features in the image, characterize high-resolution heterogeneity, and quantify the frequency of co-occurring adjacent voxel pairs with the same grayscale intensity in a specified direction [81].
3) 16 gray level run-length matrix (GLRLM)-based features: describe coarse texture features in the image, characterize low-resolution heterogeneity, and quantify the distribution of consecutively occurring intensity values of the same gray level in a specified direction [82].

The detailed list of 56 radiomic features was provided in Table 1.

Feature selection was subsequently implemented to remove the redundant features and prevent the potential overfitting. Following previous radiomic feature selection studies [83, 84], the process of feature selection in this study encompassed three steps:

1) Pearson correlation analysis [85] was performed on radiomic maps at each time $t$ to identify inter-feature correlations, which yielded a correlation matrix for each time step $t$. The average correlation matrix was derived by averaging all obtained correlation matrices.
2) Hierarchical clustering was applied to the average correlation matrix to group features into well-separated clusters based on distance [86]. Similar or highly correlated features were positioned closer in the resulting dendrogram. A total of $\widetilde{N}$ clusters were determined based on a specific distance threshold (i.e., cut-off value); features whose distances were below this threshold were considered as one cluster.
3) Spearman correlation [87] was subsequently used to select representative features. Within each cluster, the feature that demonstrated the highest Spearman correlation with the measured ventilation was selected to form $\widetilde{N}$ representative features for following analysis.

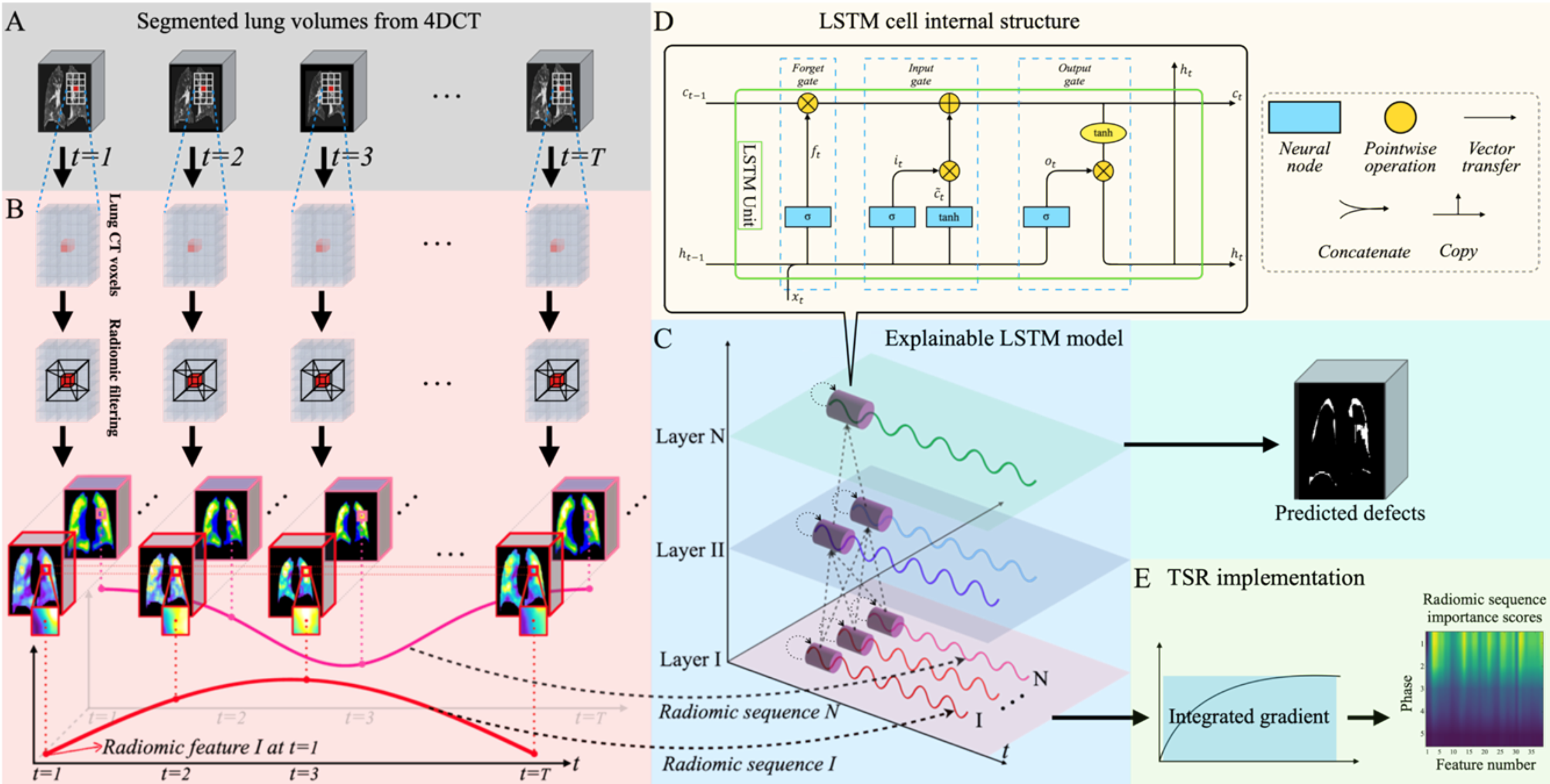

*__Fig. 1.__ The overall design of the explainable radiomic sequence model. (A) Segmented lung volumes from 4DCT images. (B) The evolution of the radiomic feature at each voxel coordinate throughout the respiratory cycle can be modelled as a radiomic sequence (e.g., as presented by red wave). (C) Explainable LSTM model implementation for ventilation defects identification. The purple cylinders represent the LSTM cells in recurrent layers. (D) The internal structure of the LSTM cell. (E) TSR implementation for identifying key radiomic sequences and key time steps.*

*Table 1*
*Fifty-six radiomic features included in this study.*

| | | | | | |
|---|---|---|---|---|---|
| Intensity-based features | 1 | Mean | Gray level co-occurrence matrix (GLCOM)-based features | 29 | Inverse Difference |
| | 2 | Variance | | 30 | Inverse Difference Moment |
| | 3 | Skewness | | 31 | Info Measure Correlation 1 |
| | 4 | Intensity histogram kurtosis | | 32 | Info Measure Correlation 2 |
| | 5 | Median | | 33 | Inverse Difference Moment Normalized |
| | 6 | Minimum grey level | | 34 | Inverse Difference Normalized |
| | 7 | 10th percentile | | 35 | Inverse Variance |
| | 8 | 90th percentile | | 36 | Joint maximum |
| | 9 | Maximum grey level | | 37 | Sum Average |
| | 10 | Interquartile range | | 38 | Sum Entropy |
| | 11 | Range | | 39 | Sum Variance |
| | 12 | Mean absolute deviation | | 40 | Joint Variance |
| | 13 | Robust mean absolute deviation | Gray level run-length matrix (GLRLM)-based features | 41 | Short Run Emphasis |
| | 14 | Median absolute deviation | | 42 | Long Run Emphasis |
| | 15 | Coefficient of variation | | 43 | Gray Level Non-Uniformity |
| | 16 | Quartile coefficient of dispersion | | 44 | Gray Level Non-Uniformity Normalized |
| | 17 | Energy | | 45 | Run Length Non-Uniformity |
| | 18 | Root mean square | | 46 | Run Length Non-Uniformity Normalized |
| Gray level co-occurrence matrix (GLCOM)-based features | 19 | Auto Correlation | | 47 | Run Percentage |
| | 20 | Cluster Prominence | | 48 | Low Gray Level Run Emphasis |
| | 21 | Cluster Shade | | 49 | High Gray Level Run Emphasis |
| | 22 | Cluster Tendency | | 50 | Short Run Low Gray Level Emphasis |
| | 23 | Contrast | | 51 | Short Run High Gray Level Emphasis |
| | 24 | Correlation | | 52 | Long Run Low Gray Level Emphasis |
| | 25 | Differential Entropy | | 53 | Long Run High Gray Level Emphasis |
| | 26 | Dissimilarity | | 54 | Gray Level Variance |
| | 27 | Joint Energy | | 55 | Run Length Variance |
| | 28 | Joint Entropy | | 56 | Run Entropy |

### 3.2. Explainable LSTM Model Design

A specially designed LSTM model was developed to associate the radiomic sequences $\boldsymbol{\Phi}$ with measured ventilation defects, as shown in Fig. 1(C). The developed LSTM model consisted of 5 recurrent layers with decreasing numbers of LSTM cells: 128, 64, 32, 16, and 8, respectively. LSTM cell (Fig. 1D) consists of three gates: the input gate, $i_t$, the forget gate, $f_t$, and the output gate, $o_t$. The hidden state $h_t$ of an LSTM cell at time step $t$ is updated by integration of the input $\theta_t$, input gate $i_t$, forget gate $f_t$, output gate $o_t$, cell state $c_t$, and hidden state $h_{t-1}$ at preceding time step $t-1$ [67, 88]. As shown in the Fig. 1(D), The forget gate $f_t$ modulates the retention of the previous cell state $c_{t-1}$, determining how much of the past information is carried forward. The input gate $i_t$ governs the extent to which this new candidate information is incorporated into the cell state $c_t$. The output gate $o_t$ dictates how much of the cell state $c_t$ is used to compute the hidden state $h_t$ through an element-wise product with the hyperbolic tangent of $c_t$. In this process, the LSTM cell selectively remember or forget information, updates the hidden state to control how information flows in and out of the internal states of the network [89, 90]. In the first recurrent layer, a series of hidden states (i.e., hidden state sequence) was derived, which can be denoted as $\boldsymbol{H} = \{h_1, h_2, \cdots, h_t\}$. The recurrent layer learns characteristics of radiomic sequences from different aspects at each time step $t$, thereby capturing the time dependencies [88]. The subsequent 4 recurrent layers adhered to a similar design, and the input of each layer was the hidden state sequences $\boldsymbol{H}$ derived from the previous layer. A dense layer with a sigmoid activation function was finally employed to generate the binary classification prediction (i.e., lung defects or healthy lung).

The Temporal Saliency Rescaling (TSR) technique was subsequently employed to explain key radiomic sequences and key time steps in the ventilation defects identification, which is built upon the integrated gradient (IG) technique [91], as shown in Fig. 1(E). Let function $L : \mathbb{R}^{\widetilde{N} \times T} \rightarrow [0,1]$ represents our neural network, $\overline{\boldsymbol{\Phi}} \in \mathbb{R}^{\widetilde{N} \times T}$ be the baseline input (i.e., a zero-embedding matrix considered as a non-informative reference point). Consider the straight-line path from the baseline $\overline{\boldsymbol{\Phi}}$ to the input $\boldsymbol{\Phi}$ and compute the gradients at all points along the path. IG is obtained by cumulating these gradients and is defined as the following equation:

$$IG(\boldsymbol{\Phi}) = \left(\boldsymbol{\Phi} - \overline{\boldsymbol{\Phi}}\right) \cdot \int_{\alpha=0}^{1} \frac{\partial L\left(\overline{\boldsymbol{\Phi}} + \alpha\left(\boldsymbol{\Phi} - \overline{\boldsymbol{\Phi}}\right)\right)}{\partial \boldsymbol{\Phi}} d\alpha \tag{4}$$

where $\alpha$ is the interpolation parameter. IG quantifies the cumulative contribution of the input in the model prediction from the baseline state $\overline{\boldsymbol{\Phi}}$ to the actual state $\boldsymbol{\Phi}$ [91]. Based on the IG technique, TSR decouples the feature importance analysis into assessments of time relevance scores $\boldsymbol{S}^{time}$ and feature relevance scores $\boldsymbol{S}^{feature}$ to identify the key radiomic sequences [77]. Time relevance scores $\boldsymbol{S}^{time}$ were assessed by observing the IG changes when specific 4DCT time step was omitted. Specifically, to calculate the time relevance score at time step $t$, $\boldsymbol{\Phi}^t$ is defined by setting $\theta_t = 0$ in $\boldsymbol{\Phi}$. By substituting $\boldsymbol{\Phi}$ and $\boldsymbol{\Phi}^t$ separately into Eq. (4) and calculating the difference, the time relevance score $S_t$ for time step $t$ was obtained as follows:

$$S_t = |IG(\boldsymbol{\Phi}) - IG(\boldsymbol{\Phi}^t)| \tag{5}$$

By implementing Eq. (5) on each time step, the time relevance scores $\boldsymbol{S}^{time}$ were obtained (i.e., $\boldsymbol{S}^{time} = \{S_{t=1}, S_{t=2}, \cdots, S_{t=T}\}$). Similarly, feature relevance scores $\boldsymbol{S}^{feature}$ were assessed by observing the IG changes when specific radiomic feature was omitted. Specifically, to calculate the feature relevance score of radiomic feature $n$, $\boldsymbol{\Phi}^n$ is defined by setting $\Phi_n = 0$ in $\boldsymbol{\Phi}$. By separately substituting $\boldsymbol{\Phi}$ and $\boldsymbol{\Phi}^n$ into Eq. (4) and calculating the difference, the feature relevance score $S_n^{feature}$ of radiomic feature $n$ was obtained as follows:

$$S_n = |IG(\boldsymbol{\Phi}) - IG(\boldsymbol{\Phi}^n)| \tag{6}$$

By computing Eq. (6) on each radiomic feature, the feature relevance scores $\boldsymbol{S}^{feature}$ were obtained (i.e., $\boldsymbol{S}^{feature} = \{S_{n=1}, S_{n=2}, \cdots, S_{n=\tilde{N}}\}$). The temporal saliency map can be derived by taking the outer product of time relevance scores $\boldsymbol{S}^{time}$ and feature relevance scores $\boldsymbol{S}^{feature}$:

$$Temporal\ saliency\ map = \boldsymbol{S}^{feature} \cdot (\boldsymbol{S}^{time})^T \tag{7}$$

The horizontal axis of the map represents the feature index from 1 to $\tilde{N}$, the vertical axis of the map represents the time step from 1 to $T$. The element at coordinate $(n, t)$ of the temporal saliency map is the product of $S_n$ and $S_t$, and represents the quantified importance score for corresponding feature $f_n^t$. The importance score of radiomic sequence $\Phi_n$ was derived by averaging the importance scores of all features $f_n^t$ within $\Phi_n$ across all voxel samples. The key radiomic sequences were finally identified through the analysis of the radiomic sequence importance score histogram.

## 4. Experiments and results

### *4.1. Imaging dataset*

This study utilized a public lung cancer patient dataset from the Ventilation And Medical Pulmonary Image Registration Evaluation (VAMPIRE) Dataset [9]. The VAMPIRE dataset comprises paired image acquisitions of CT and reference ventilation images (RefVI), encompassing 25 individuals with Galligas-PET/CT imaging and 20 with DTPA-SPECT/CT imaging. The respiratory phases were processed from exhalation to inhalation. The 4DCT images were reconstructed into 5 and 10 respiratory phases for PET and SPECT groups, respectively. Fig. 2 provides a visual representation of the dataset. All the 4DCT images were acquired under free-breathing conditions. Reference lung ventilation images (RefVIs), i.e., PET and SPECT images, were registered to the corresponding time-averaged 4DCT. Lung masks were provided for each phase of the 4DCTs and corresponding RefVIs of each patient. Following the original VAMPIRE studies, all 4DCT images and RefVIs were subsequently resampled with $2 \times 2 \times 2$ mm$^3$ isotropic voxel size [9], and the ground truth pulmonary defects were identified as follows: (1) voxels with ventilation intensity above $\pm 4$ standard deviations of the mean intensity of overall RefVI lung voxels were removed until the threshold converged to within 1% of the last threshold [9, 92]; (2) the region with the lowest 30% of the total intensity in PET/SPECT is considered as the ground truth pulmonary defects [9, 53].

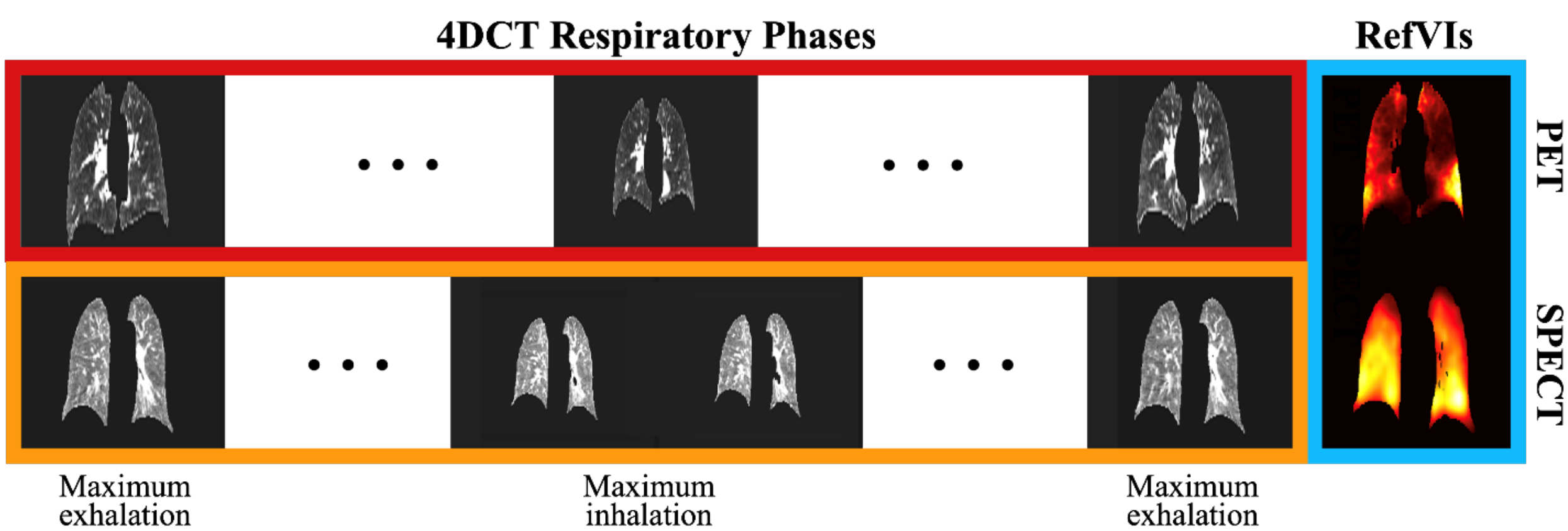

***Fig. 2.*** *Visual inspection of RefVI and 4DCT images. The upper row represents PET/CT, and the lower row represents SPECT/CT. The respiratory phases depict range from a maximum exhalation to the subsequent maximum exhalation.*

### 4.2. Ablation studies

The ablation study was conducted to rigorously identify the contributions of each component of the developed neural radiomic sequence model. Specifically, three variants of the neural radiomic sequence model were developed:

- In the first variant, LSTM with original 4DCT images was evaluated: the radiomic sequence modelling component was excluded and the raw 4DCT images served as the LSTM model input. The LSTM depicted in Fig. 1(C) was employed to identify ventilation defects by utilizing the changing voxel intensity values across respiratory phases as the sole feature input. This variant determined the impact of removing the detailed locoregional manual-defined image intensity and texture information.
- In the second variant, BiLSTM model with radiomic sequences was evaluated: the LSTM model in Fig. 1(C) was replaced by a BiLSTM. This variant assessed the model performance between LSTM and BiLSTM. Specifically, the BiLSTM model consisted of two LSTM components: a forward LSTM and a backward LSTM. The forward component processes the input sequence from beginning to end. The backward component processes the input sequence from end to beginning. In each direction, the processing of radiomic sequences was modeled as a recurrent process with its own hidden state [93]. The bidirectional hidden states were concatenated, and a dense layer with a sigmoid activation function was applied to this concatenated state to generate the prediction of ventilation defects.
- In the third variant, Transformer with radiomic sequences was evaluated: the explainable LSTM model in Fig. 1(C) was replaced by a Transformer encoder. This variant assessed the performance between LSTM and Transformer encoder. The Transformer encoder was composed of positional encoding and four identical transformer encoder blocks. A fixed sinusoidal positional encoding scheme was applied to the input before they are fed into the encoder to mark the inherent sequential order in the input data [94]. Each encoder block consisted of a multi-head self-attention layer (4 attention heads) to dynamically attend to distinct regions of the input sequence [94] and a feed-forward neural network layer to enhance discriminative feature representations. Additionally, residual connections were incorporated after each encoder block, which directly added the original input of the layer to its output. This design aimed to preserve input information and improve gradient flow [95]. The ventilation defects were predicted by adopting a dense layer with sigmoid activation.

The training settings for all variant models were identical to those of our proposed model. Each variant was evaluated and compared with the neural radiomic sequence model to assess its relative performance in identifying ventilation defects. Detailed evaluation metrics are presented in Section 4.5.

### *4.3. Comparison studies*

The comparison studies were also included to further investigate the model performance of the developed neural radiomic sequence model:

- **U-Net with original 4DCT**: the prediction model based on U-Net CNN using the original 4DCT images as input. The U-Net model in this study was composed of an encoding part and a decoding part [96]. The encoding part contained four convolutional blocks; each block contains two 3×3×3 convolutional layers, followed by a rectified linear unit (ReLU) activation and a 2×2×2 max pooling operation. The decoding part is composed of four up-convolutional blocks. Each block in this part included a 2×2×2 transposed convolution to up-sample the feature maps, a concatenation with high-resolution features from the encoding part to combine the feature and spatial information, and a convolutional block to refine the representation. The ventilation defects were predicted by adopting a 1×1×1 convolutional layer with sigmoid activation.
- **U-Net++ with original 4DCT**: the prediction model based on U-Net++ using the original 4DCT images as input. Building on the U-Net model, U-Net++ added additional skip connections between the intermediate convolutional blocks of the encoding and decoding parts [97]. Features from earlier blocks in the encoder were not only connected to their corresponding blocks in the decoder but also to multiple blocks in the decoder (e.g. features from the first block in the encoder were connected to the first, second, third, and fourth blocks in the decoder). The ventilation defects were predicted by adopting a 1×1×1 convolutional layer with sigmoid activation.
- **Res-UNet with original 4DCT**: the prediction model based on Res-UNet using the original 4DCT images as input. The Res-UNet model introduced residual connections into both the encoding and decoding parts (i.e., all convolution blocks were replaced by residual blocks) [95]. The residual block consists of two 3×3×3 convolutional layers, followed by ReLU activation functions and a 2×2×2 max pooling operation. The input to each residual block is directly added to the output of the second convolutional layer via a skip connection. The residual operation allowed the network to learn identity mappings and helped preserve information of input data [95]. The ventilation defects were predicted by adopting a 1×1×1 convolutional layer with sigmoid activation.

- **Random forest with original 4DCT**: the prediction model based on random forest (RF) using the original 4DCT images. RF is a typical machine learning model that adopts a hierarchical tree structure [98]. Each internal node represents a decision based on a specific input variable, while each leaf node provides a prediction for the output variable. For each voxel, the intensity values at different time steps served as different input features. These features were fed into the RF model to predict whether the voxel represented a ventilation defect.

All comparison models employed the same Adam optimization algorithm, early stopping strategy, and loss function as our proposed model. For U-Net with 4DCT, U-Net++ with 4DCT, and Res-UNet with 4DCT, the batch size was set to 16. For RF with 4DCT, hyperparameters were optimized using random search, and training was performed with 100 decision trees, Gini impurity criterion, and bootstrap sampling.

### *4.4. Implementation details*

The radiomic filtering was implemented according to previous lung radiomic filtering studies: (1) the intensity-based features were extracted directly from the lung volume images [79], (2) a fixed bin number (n=64) image discretization was adopted to calculate the second-order features (i.e., GLCOM-based and GLRLM-based features) [44], and (3) a $26 \times 26 \times 26$ mm$^3$ sized kernel was employed for effective regional feature extraction [38, 99]. All 56 radiomic features were averaged over 13 directions [100] to approximate rotational invariance [42, 82, 101, 102]. All the radiomic filtering calculations were performed using an in-house developed radiomics filtering toolbox with MATLAB (MATLAB R2023a; MathWorks, Natick, Ma). The toolbox has been comprehensively validated against the image biomarker standardization initiative (IBSI) standardization [103] and the digital phantoms [104]. Additionally, the toolbox has been specifically optimized for voxel-based, rotationally invariant calculations in 3D spaces [44]. For the developed LSTM model, training was carried out for up to 500 epochs using the Adam optimization algorithm with a learning rate of 0.0001 and a gradient clipping threshold of 1.0. Early stopping was implemented to mitigate overfitting. The binary cross-entropy loss function was adopted, and a batch size of 2048 was maintained throughout the training process.

The hierarchical clustering of radiomic filtering maps was evaluated using the Cophenetic Correlation Coefficient (CCC) [105]. A CCC value above 0.75 indicates that the dendrogram produced by the hierarchical clustering accurately represents feature distances [83]. The optimal cut-off value was determined by the highest silhouette coefficient (SC) value [106], where an SC value above 0.7 indicates better cluster compactness and separation [107].

To rigorously assess comparison model performance, five-fold cross-validation was conducted on 25 PET and 20 SPECT cases. In each fold, 80% of the data were used for training and the remaining 20% for testing, ensuring that all cases were included in the test set exactly once. Evaluation metrics included Dice similarity coefficient, area under the ROC curve (AUC-ROC), sensitivity, and classification accuracy. Statistical significance across different models was determined using the Wilcoxon signed-rank test (two-tailed, significance level = 0.01) for all performance metrics.

All the calculations were carried out in computational workstation with 16 Core Intel Core i7-13700KF CPU @ 3.4 GHz, 128GB DDR4 memory (4 × 32GB @ 3200 MHz), and Nvidia GeForce RTX 4070 Card. The TSR technique was implemented with TSInterpret Library [108].

## *4.5. Results of feature selection*

Fig. 3 exhibits the feature correlation heatmap after hierarchical clustering. The heatmap visualizes the correlations between different features using color gradations. The intensity of the color reflects the strength of the correlation, ranging from 1 (indicating positive correlation, shown in blue) to -1 (indicating negative correlation, shown in red). The hierarchical clustering dendrogram illustrates the clustering relationships among features, with a CCC value of 0.80 (>0.75). Features with higher correlation were positioned closer together in the dendrogram. The cut-off value (=0.02) with the highest SC value of 0.79 (>0.7) was utilized to categorize 38 feature clusters. In each cluster, the feature exhibiting the highest Spearman correlation with the ground truth was finally selected, thereby 38 features were finally selected.

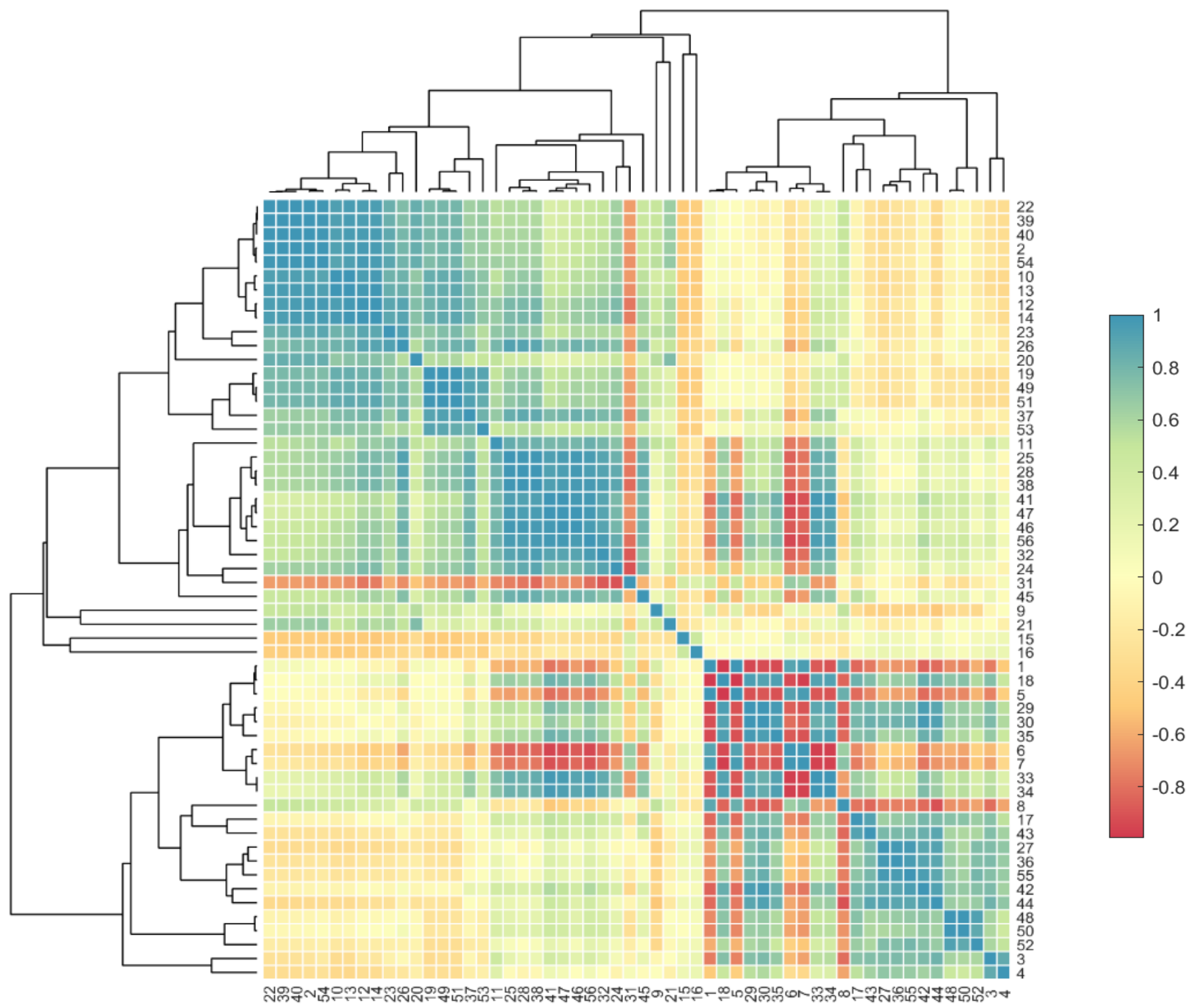

***Fig. 3.*** *The feature correlation heatmap after hierarchical clustering. The number of x or y axis represents the feature number. The intensity of the color indicates the strength of the correlation between two features, ranging from 1 (indicating positive correlation, shown in blue) to -1 (indicating negative correlation, shown in red). The dendrogram on the left side or top of the heatmap display the clustering relationship between features. Features that were more highly correlated were positioned closer together in the dendrogram.*

### 4.6. *Results of ventilation defects prediction*

Figs. 4 through 9 present representative examples of binary ventilation defect predictions generated by the proposed neural radiomic sequence model and benchmarked against multiple comparative models. Each figure consists of three panels: Part A of these figures showed original RefVI and the identified lung defects regions by thresholding the lowest 30% of the total intensity of the RefVI. Part B presented the corresponding 4DCT images from maximum exhalation to the subsequent maximum exhalation. Part C exhibited the predicted defects from all comparative models. Figs. 4 to 6 correspond to cases validated using SPECT imaging, while Figs. 7 to 9 show PET-based examples.

Visually, the proposed model demonstrated strong agreement with the ground truth across both modalities. In Example I (Fig. 4), it accurately identified ventilation defects in the upper lung region. Example II (Fig. 5) showed precise localization in the mid-right and lower-left zones. In Example III (Fig. 6), the model captured upper-right lung defects, including unlabeled areas in the RefVI. In Examples IV to VI (Figs. 7 to 9), the predictions also closely matched the reference, particularly within the mid-left and lower-right regions, respectively. These findings highlight the robustness of the proposed method in detecting both annotated and physiologically plausible unannotated ventilation abnormalities. Notably, the BiLSTM and Transformer models trained on radiomic sequences produced similarly consistent visual effects, underscoring the benefits of temporal modeling in enhancing regional defect localization.

In contrast, models trained on raw 4DCT data yielded inferior visual consistency and spatial precision. The LSTM model demonstrated partial capability in Examples I through IV but consistently overestimated the extent of the defect regions and failed to detect abnormalities in Examples V and VI. The RF model produced scattered and noisy outputs (see Figs. 4 and 5) or even failed entirely in defects detection (see Figs. 6 to 9). Both U-Net and U-Net++ exhibited limited detection ability: U-Net with original 4DCT failed in most cases (Examples I, II, III, and V) and U-Net++ occasionally localizing parts of the defect regions but often significantly overestimating boundaries (see Figs. 4, 6, 7, and 9). Res-UNet with original 4DCT produced mixed results, underestimating defect areas in some cases (see Example II and V) and overestimating in others (see example IV and VI), with only marginally accurate prediction in Example III.

Quantitative results derived from five-fold cross-validation are summarized in Table 2, and the corresponding ROC curves are shown in Fig. 10. The proposed neural radiomic sequence model achieved the highest overall performance across all evaluation metrics. For 25 PET cases, it achieved a mean Dice score of 0.78, an AUC-ROC of 0.85, sensitivity of 0.78, and accuracy of 0.76. For 20 SPECT cases, the results were similarly robust, with a Dice of 0.78, AUC-ROC of 0.84, sensitivity of 0.78, and accuracy of 0.74. The BiLSTM and Transformer models with radiomic sequences also showed strong performance, with BiLSTM achieving Dice scores of 0.78 for both PET and SPECT, AUC-ROC values of 0.85 and 0.84, sensitivity of 0.77, and accuracy of 0.76/0.74, respectively. The Transformer model attained Dice scores of 0.77 for both modalities, AUC-ROC of 0.85/0.84, sensitivity of 0.76/0.74, and accuracy of 0.77/0.75.

In comparison, models based on original 4DCT showed consistently lower performance. U-Net achieved mean Dice scores of 0.51 in both PET and SPECT cases, with AUC-ROC of 0.68/0.65, sensitivity of 0.40/0.47, and accuracy of 0.67/0.62. U-Net++ improved marginally with Dice scores of 0.62/0.57, AUC-ROC of 0.72/0.69, sensitivity of 0.64/0.63, and accuracy of 0.67/0.60. Res-UNet yielded Dice of 0.59/0.67, AUC-ROC of 0.73/0.74, sensitivity of 0.57/0.68, and accuracy of 0.68 for both. RF achieved Dice scores of 0.58/0.52, AUC-ROC of 0.63/0.67, sensitivity of 0.59/0.48, and accuracy of 0.60/0.63. The LSTM model performed moderately better, with Dice score of 0.69/0.66, AUC-ROC of 0.77/0.71, sensitivity of 0.65 for both, and accuracy of 0.70/0.63.

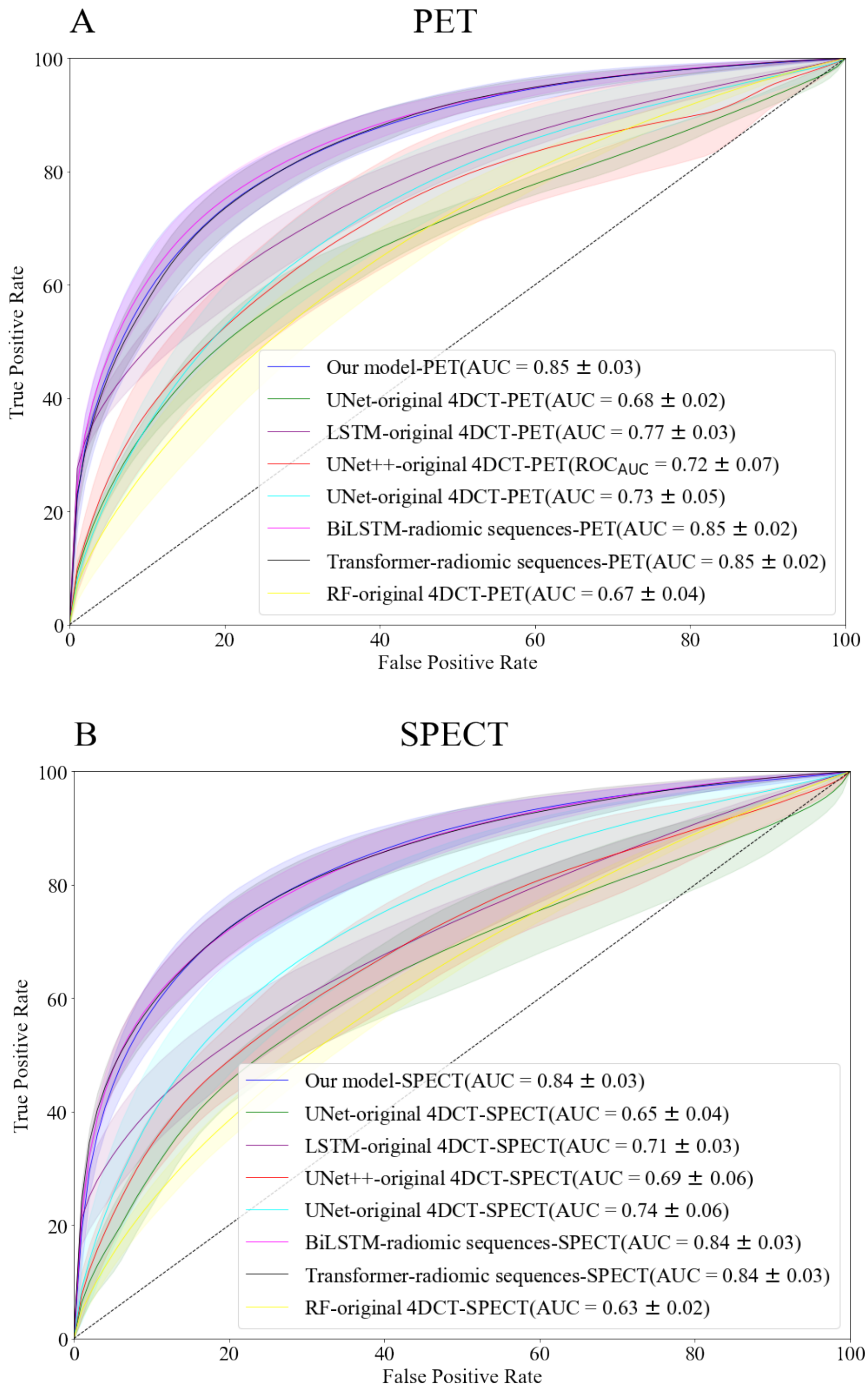

***Fig. 10.*** *ROC Curves from our proposed model (blue line), U-Net with original 4DCT (green line), LSTM with original 4DCT (purple line), U-Net++ with original 4DCT (red line), Res-UNet with original 4DCT (cyan line), BiLSTM with radiomic sequences (magenta line), Transformer with radiomic sequences (black line), and RF with original 4DCT (yellow line) in (A) PET cases, and (B) SPECT cases.*

***Table 2*** *Five-fold cross-validation ventilation quantification results in all comparative studies.*

| | Models | Dice | AUC-ROC | Sensitivity | Accuracy |
|---|---|---|---|---|---|
| **PET** | **Our model** | 0.78(0.74-0.79) | 0.85(0.80-0.88) | 0.78(0.72-0.83) | 0.76(0.72-0.79) |
| | **BiLSTM with radiomic sequences** | 0.78(0.74-0.79) | 0.85(0.81-0.87) | 0.77(0.72-0.82) | 0.76(0.72-0.78) |
| | **Transformer with radiomic sequences** | 0.77(0.73-0.79) | 0.85(0.81-0.87) | 0.76(0.73-0.81) | 0.77(0.73-0.79) |
| | **U-Net with original 4DCT** | 0.51(0.44-0.55)* | 0.68(0.65-0.70)* | 0.40(0.32-0.44)* | 0.67(0.62-0.70)* |
| | **U-Net++ with original 4DCT** | 0.62(0.55-0.75)* | 0.72(0.61-0.80)* | 0.64(0.50-0.80)* | 0.67(0.58-0.73)* |
| | **Res-UNet with original 4DCT** | 0.59(0.36-0.69)* | 0.73(0.67-0.80)* | 0.57(0.24-0.73)* | 0.68(0.65-0.74)* |
| | **RF with original 4DCT** | 0.58(0.48-0.64)* | 0.63(0.60-0.65)* | 0.59(0.51-0.74)* | 0.60(0.57-0.64)* |
| | **LSTM with original 4DCT** | 0.69(0.65-0.74)* | 0.77(0.73-0.80)* | 0.65(0.56-0.71)* | 0.70(0.66-0.74)* |
| **SPECT** | **Models** | **Dice** | **AUC-ROC** | **Sensitivity** | **Accuracy** |
| | **Our model** | 0.78(0.74-0.82) | 0.84(0.80-0.87) | 0.78(0.68-0.86) | 0.74(0.72-0.79) |
| | **BiLSTM with radiomic sequences** | 0.78(0.73-0.82) | 0.84(0.80-0.86) | 0.77(0.67-0.84) | 0.74(0.71-0.78) |
| | **Transformer with radiomic sequences** | 0.77(0.73-0.82) | 0.84(0.80-0.86) | 0.74(0.64-0.80) | 0.75(0.70-0.78) |
| | **U-Net with original 4DCT** | 0.51(0.40-0.58)* | 0.65(0.61-0.70)* | 0.47(0.26-0.72)* | 0.62(0.58-0.67)* |
| | **U-Net++ with original 4DCT** | 0.57(0.28-0.70)* | 0.69(0.59-0.75)* | 0.63(0.17-0.88)* | 0.60(0.49-0.69)* |
| | **Res-UNet with original 4DCT** | 0.67(0.64-0.69)* | 0.74(0.67-0.85)* | 0.68(0.58-0.78)* | 0.68(0.62-0.69)* |
| | **RF with original 4DCT** | 0.52(0.48-0.57)* | 0.67(0.60-0.70)* | 0.48(0.42-0.54)* | 0.63(0.59-0.67)* |
| | **LSTM with original 4DCT** | 0.66(0.60-0.70)* | 0.71(0.66-0.77)* | 0.65(0.50-0.82)* | 0.63(0.58-0.72)* |

*"*" indicates a statistically significant difference compared with results of proposed model.*

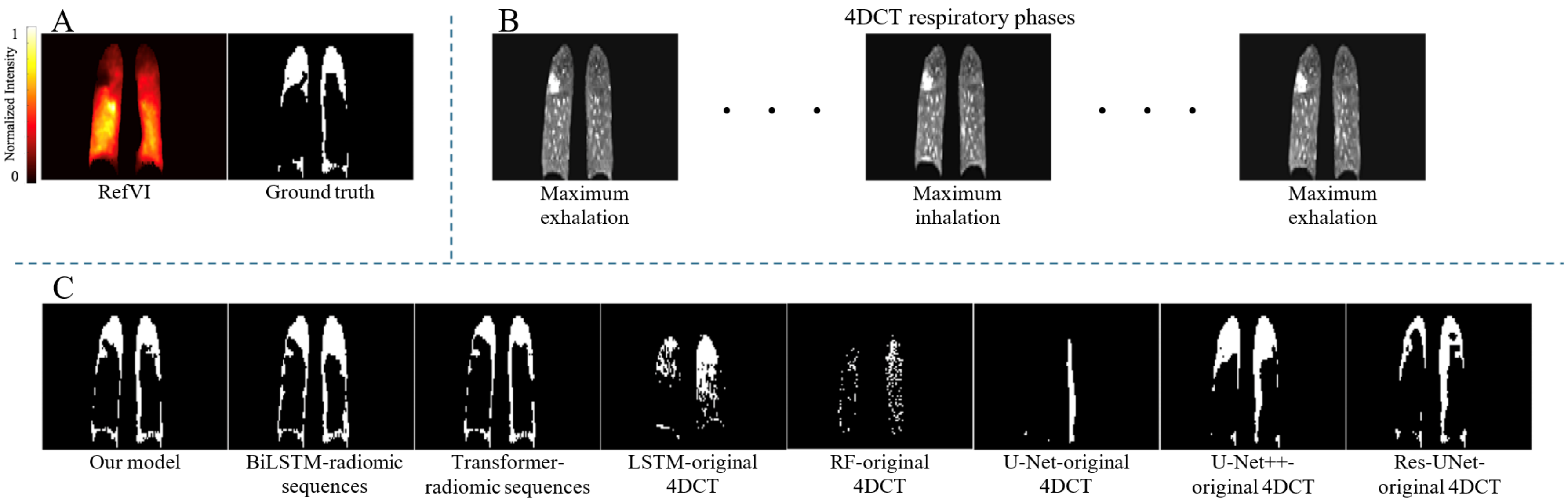

***Fig. 4.*** *Example I of PET, 4DCT, and model prediction demonstration. (A) original RefVI and identified lung defects regions by thresholding the lowest 30% of the total intensity of the RefVI; (B) corresponding 4DCT from a maximum exhalation to the subsequent maximum exhalation; (C) predicted defects regions from all comparative models.*

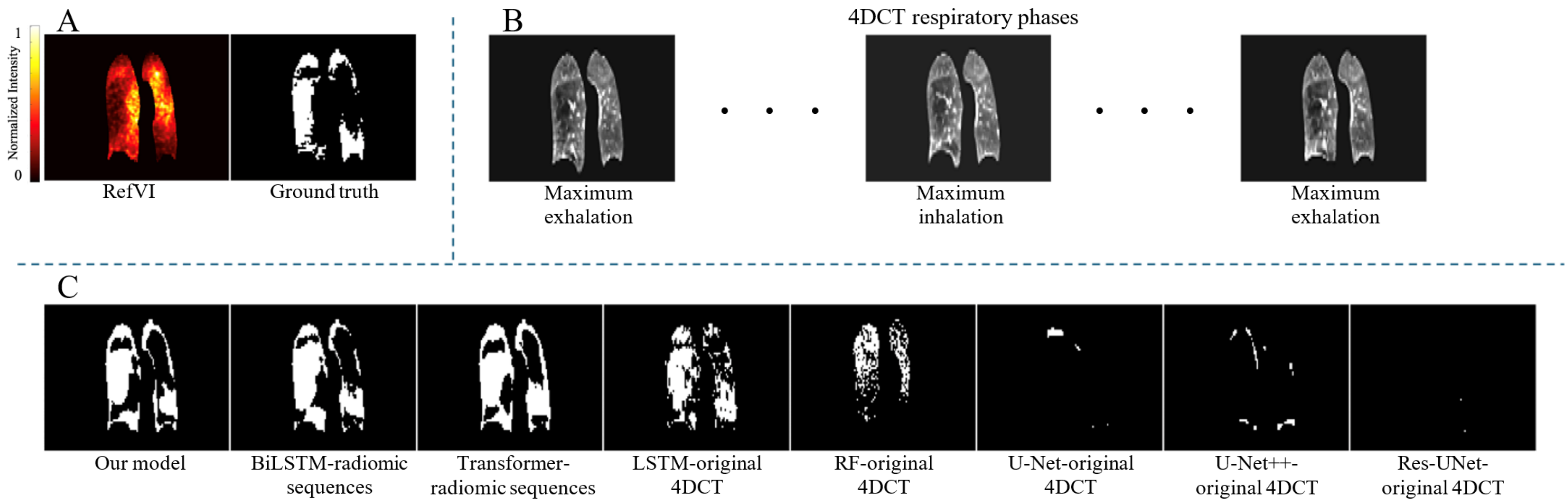

***Fig. 5.*** *Example II of PET, 4DCT, and model prediction demonstration. (A) original RefVI and identified lung defects regions by thresholding the lowest 30% of the total intensity of the RefVI; (B) corresponding 4DCT from a maximum exhalation to the subsequent maximum exhalation; (C) predicted defects regions from all comparative models.*

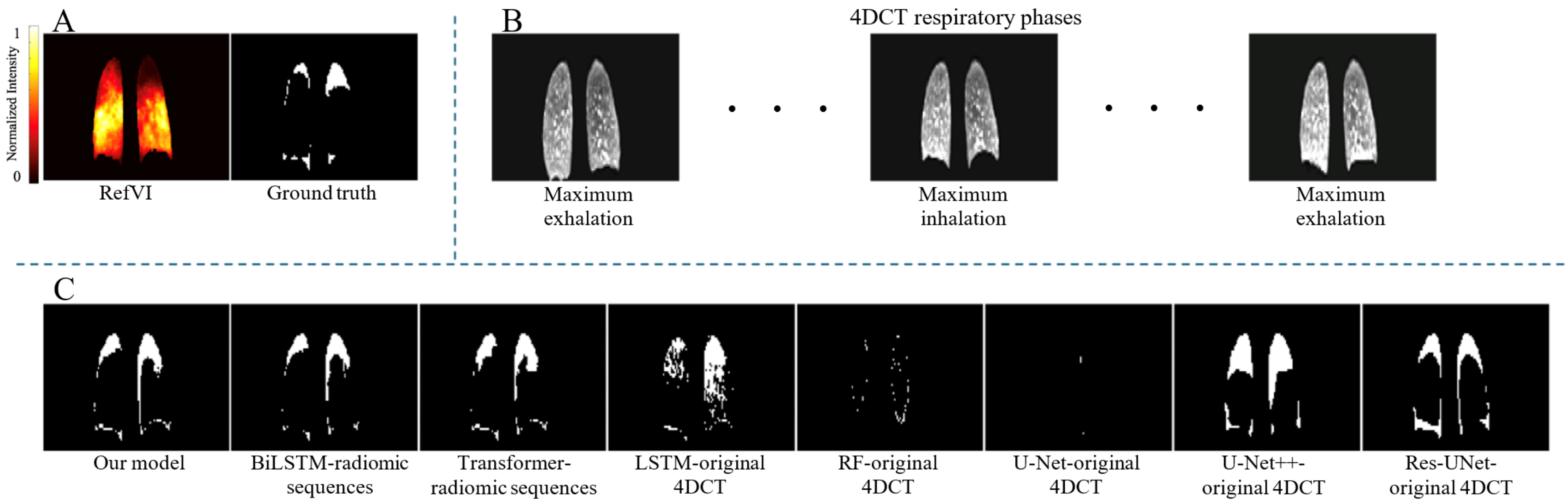

***Fig. 6.*** *Example III of PET, 4DCT, and model prediction demonstration. (A) original RefVI and identified lung defects regions by thresholding the lowest 30% of the total intensity of the RefVI; (B) corresponding 4DCT from a maximum exhalation to the subsequent maximum exhalation; (C) predicted defects regions from all comparative models.*

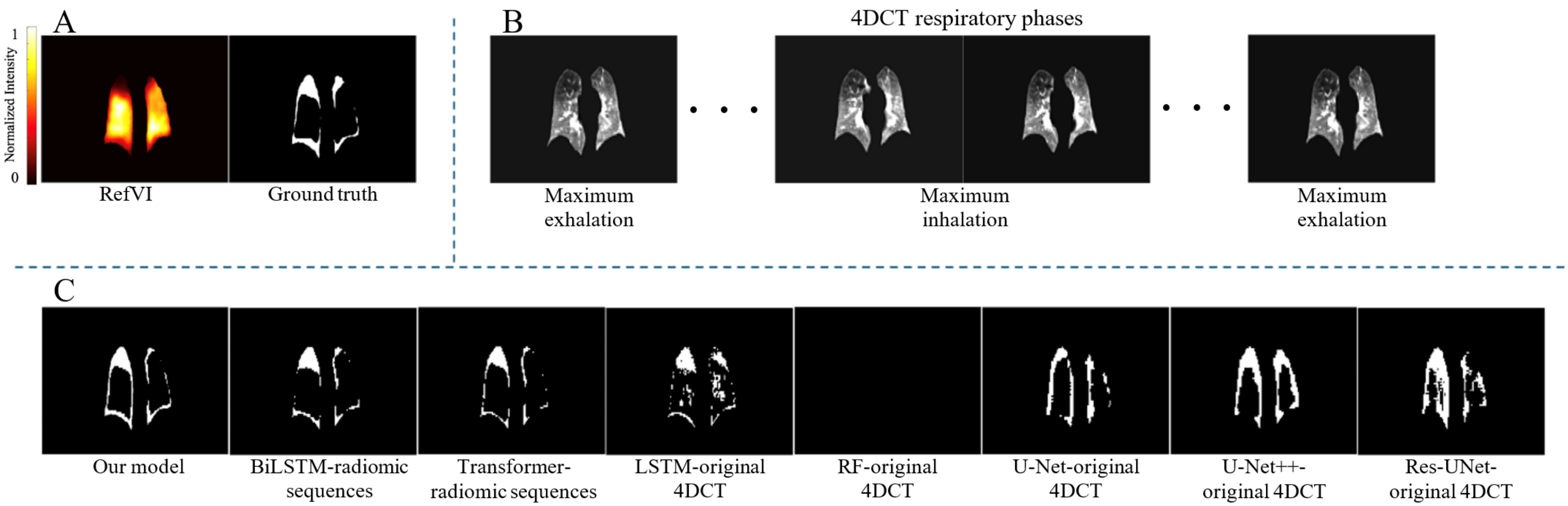

***Fig. 7.*** *Example I of SPECT, 4DCT, and model prediction demonstration. (A) original RefVI and identified lung defects regions by thresholding the lowest 30% of the total intensity of the RefVI; (B) corresponding 4DCT from a maximum exhalation to the subsequent maximum exhalation; (C) predicted defects regions from all comparative models.*

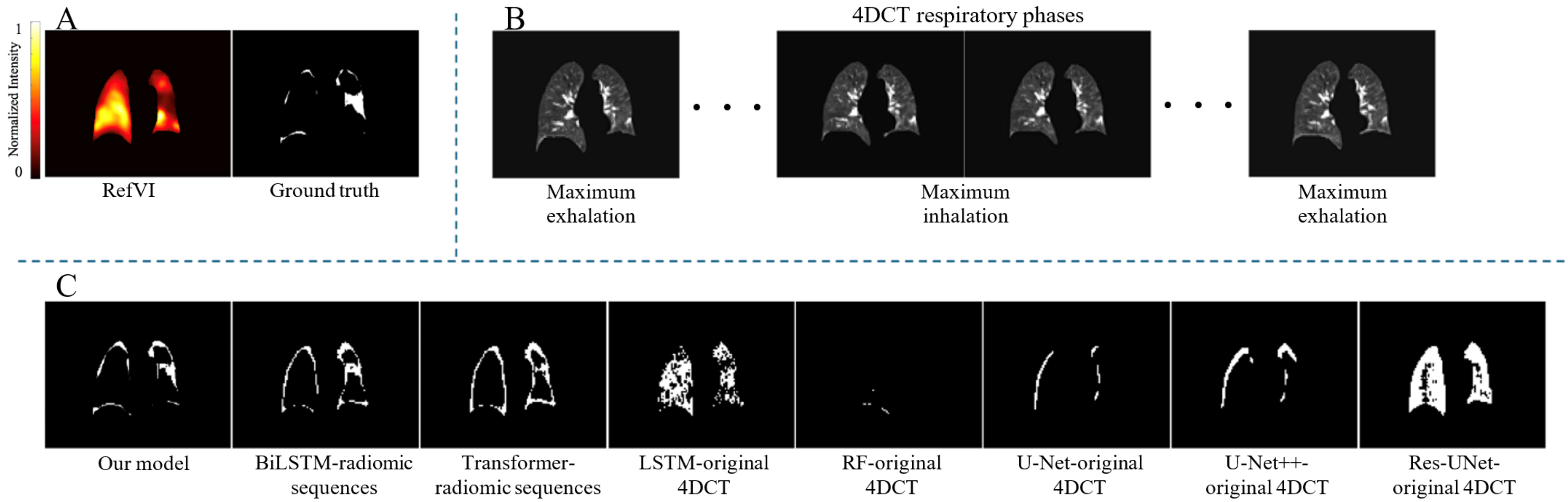

***Fig. 8.*** *Example II of SPECT, 4DCT, and model prediction demonstration. (A) original RefVI and identified lung defects regions by thresholding the lowest 30% of the total intensity of the RefVI; (B) corresponding 4DCT from a maximum exhalation to the subsequent maximum exhalation; (C) predicted defects regions from all comparative models.*

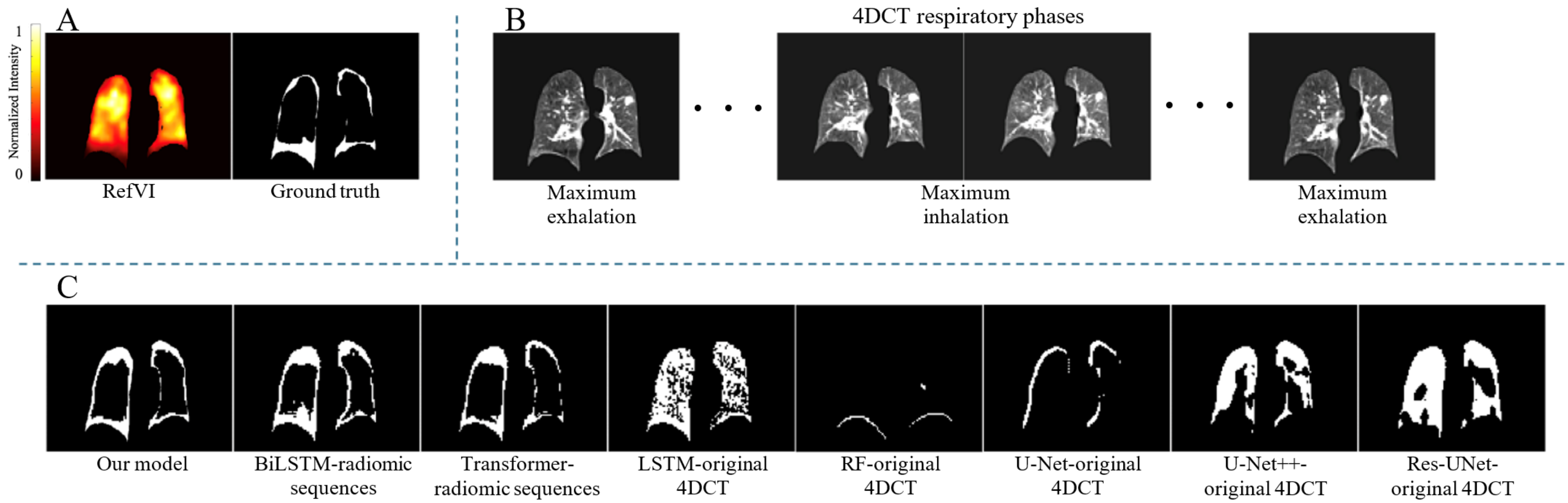

***Fig. 9.*** *Example III of SPECT, 4DCT, and model prediction demonstration. (A) original RefVI and identified lung defects regions by thresholding the lowest 30% of the total intensity of the RefVI; (B) corresponding 4DCT from a maximum exhalation to the subsequent maximum exhalation; (C) predicted defects regions from all comparative models.*

*4.7. Results of key radiomic sequences and key time steps explanation*

The average temporal saliency maps derived from PET and SPECT cohorts are presented in Fig. 11. In these maps, the horizontal axis denotes the index of radiomic sequence features, while the vertical axis corresponds to the respiratory phases, ordered sequentially as maximum exhalation, maximum inhalation, and subsequent maximum exhalation. The color intensity within the map reflects the importance of each radiomic feature at each respiratory phase, with brighter colors indicating higher importance scores. Notably, the exhalation phases appear consistently brighter than the inhalation phase across both modalities, suggesting that radiomic features extracted during exhalation are more critical for ventilation defect prediction than those derived during inhalation.

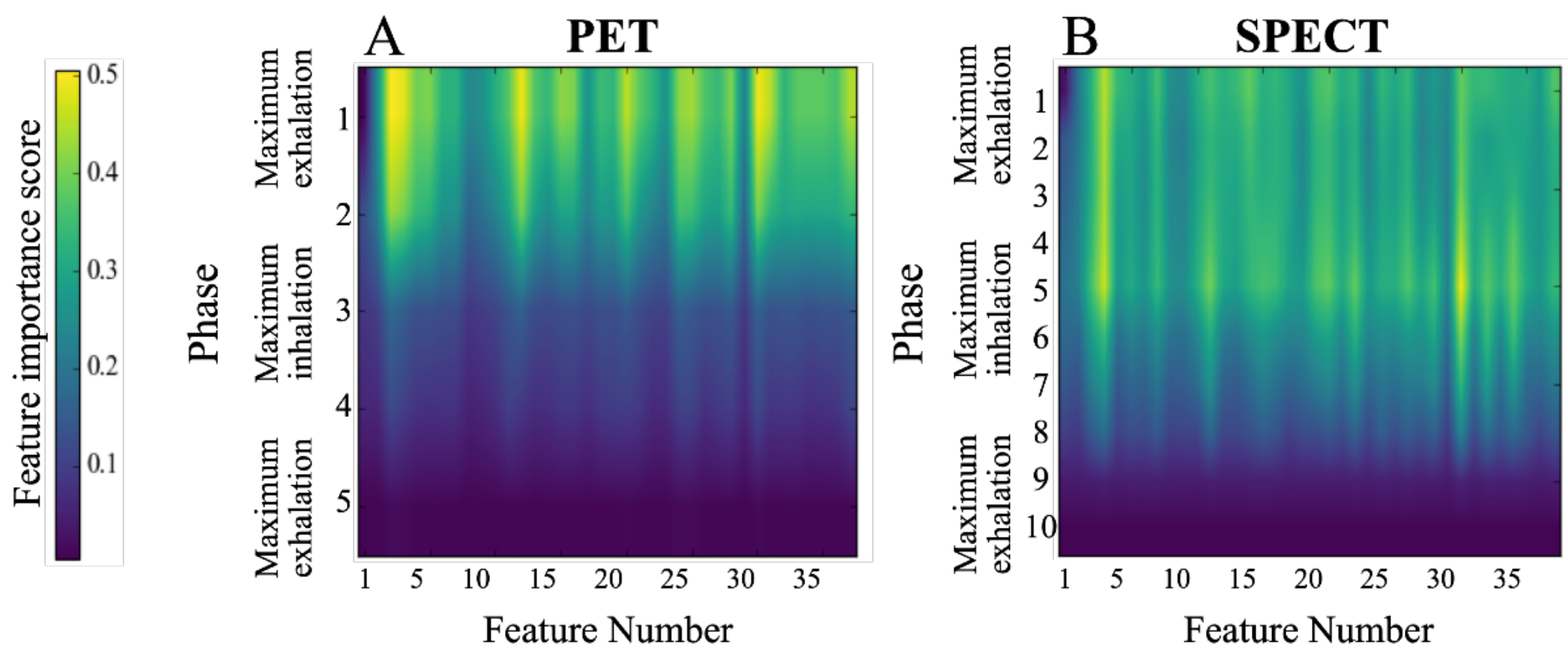

***Fig. 11.*** *Average TSMs from (A) PET cases, and (B) SPECT cases. The horizontal axis of the TSM represents radiomic sequence index, and the vertical axis represents the respiratory phase. The color in the TSM corresponds to the importance score; more intense colors indicate lower feature importance and brighter colors indicate higher feature importance.*

The histogram illustrating the distribution of radiomic sequence importance scores is shown in Fig. 12. The x-axis represents the computed importance scores, and the y-axis indicates the number of radiomic sequences within each score interval. Among all features, three radiomic sequences—namely, *Intensity-based 10th percentile (#7)*, *Intensity-based 90th percentile (#8), and GLRLM-based Run-Length Non-Uniformity (#43)*—were consistently ranked among the most predictive for identifying compromised ventilation.

The temporal variation of these three key radiomic sequences across respiratory phases is illustrated in Figure 13. Red curves represent the average trajectories for voxels located in compromised lung regions, while blue curves correspond to healthy lung voxels. The left and right panels of the figure display results for the PET (Fig. 13A) and SPECT (Fig. 13B) cohorts, respectively. All three features demonstrate a generally symmetrical pattern across the modeled respiratory cycle—spanning from maximum exhalation to inhalation and returning to exhalation—suggesting that the temporal dynamics of local intensity and texture features during inhalation largely mirror exhalation process in reverse. Despite inter-subject variability, consistent and distinct trends are observed between healthy and compromised lung regions across both modalities. In healthy lung tissue, the *Intensity-based 10th percentile* and the *GLRLM-based Run-Length Non-Uniformity* remain relatively stable throughout the respiratory cycle. The Intensity-based 90th percentile shows a mild increase during exhalation followed by a decrease during inhalation, reflecting expected physiological dynamics. In contrast, compromised regions exhibit distinct temporal dynamics. The *Intensity-based 10th percentile* shows an initial increase during exhalation, followed by a decline during inhalation. The *Intensity-based 90th percentile* also exhibits a rising trend in the exhalation phase, though the pattern is less pronounced. The *GLRLM-based Run-Length Non-Uniformity* shows an inverse trend—decreasing during exhalation and increasing during inhalation. These results suggest that compromised pulmonary regions are characterized by increased voxel intensity and reduced homogeneity during exhalation, in contrast to the more stable dynamics observed in healthy lung tissue. Furthermore, the magnitude of feature values further differentiates the two tissue types. Both *Intensity-based 10th and 90th percentile*s are consistently higher in compromised regions compared to healthy regions, indicating elevated attenuation values associated with impaired ventilation. Conversely, the *GLRLM-based Run-Length Non-Uniformity* is systematically lower in compromised regions.

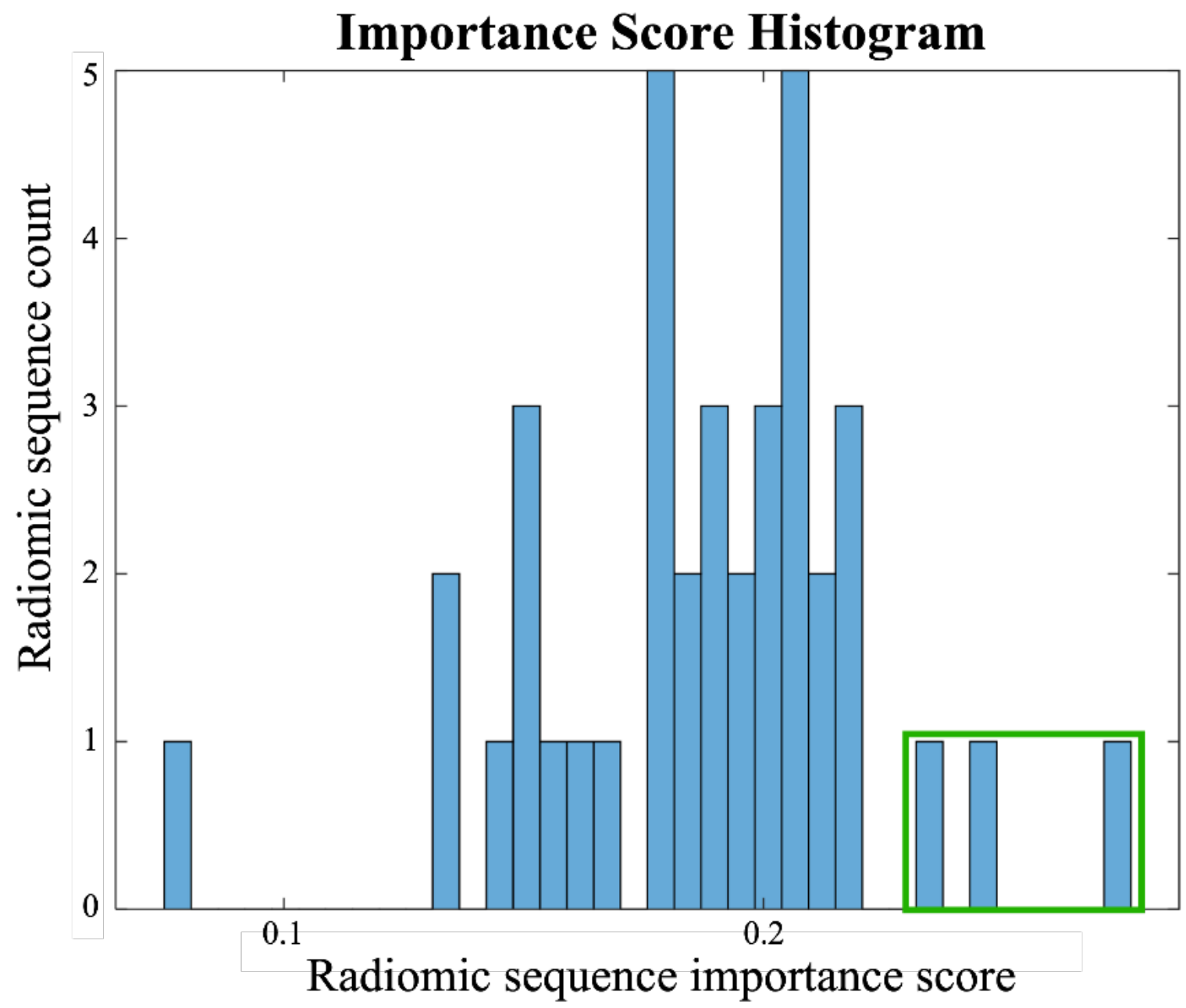

***Fig. 12.*** *The histogram of radiomic sequence importance scores. The abscissa denotes the importance scores, while the ordinate indicates the count of radiomic sequences. The green box highlights the radiomic sequences with significantly higher importance.*

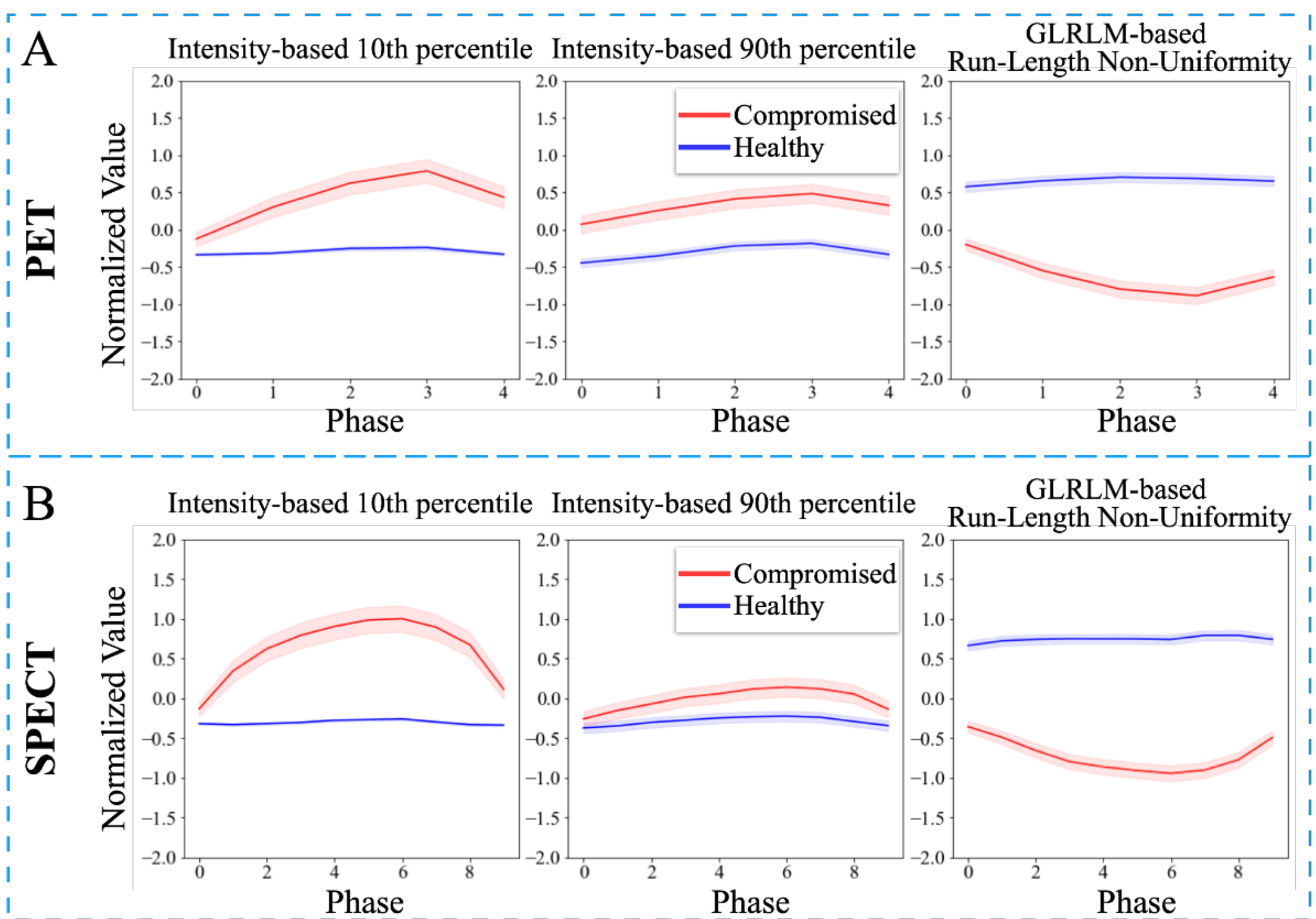

***Fig. 13.*** *Trends of key radiomic feature sequences in (A) PET cases, and (B) SPECT cases. Red line represents sequences in compromised region, and blue line represents those in healthy region.*

## 5. Discussion

This study introduces an explainable neural radiomic sequence model with spatiotemporal continuity for identifying regions of compromised pulmonary ventilation in 4DCT imaging. Traditional radiomics approaches typically derive radiomic features from averaged 4DCT images to encode static lung intensity and texture. While such methods have successfully correlated radiomic maps with ventilation maps, they often neglect the underlying motion and deformation that are central to ventilation physiology. To address this limitation, we developed a novel framework that extends radiomics from static to time-resolved analysis. At each voxel, spatially localized intensity and texture features were extracted across all respiratory phases to construct radiomic sequences, hypothesized to encode locoregional spatiotemporal heterogeneity throughout the respiratory cycle. An explainable LSTM-based network was then designed to model temporal dependencies within these sequences and detect ventilation abnormalities. The network further integrates a TSR mechanism to generate explainable attribution maps, which highlight the most discriminative radiomic features and respiratory phases driving the model's predictions.

The proposed model demonstrated strong ventilation performance on the VAMPIRE dataset, achieving a mean Dice score of 0.78 on both PET and SPECT cohorts (21 and 25 cases, respectively). Ablation studies revealed that LSTM models trained directly on raw 4DCT (without radiomic sequence modeling) yielded significantly lower performance (Dice = 0.69 for PET, 0.66 for SPECT), underscoring the critical role of radiomic sequence modeling in capturing texture and intensity heterogeneity relevant to functional impairment. Additional comparisons with BiLSTM (Dice = 0.78/0.78) and Transformer-based models (Dice = 0.77/0.77) using the same radiomic sequence input revealed no substantial performance gain, suggesting that the unidirectional LSTM architecture sufficiently models the temporal progression of radiomic features in the current dataset. Although BiLSTM can capture long-range bidirectional dependencies, its advantage is mitigated by the approximately symmetric nature of respiratory motion (i.e., ranging from exhalation to inhalation and back) [109]. Similarly, the Transformer's attention-based global modeling did not yield improvements, likely due to the limited number of temporal phases in 4DCT (5 for PET, 10 for SPECT), which constrain the benefit of long-range temporal modeling [71]. Furthermore, the increased computational demands of these more complex architectures may hinder convergence efficiency without significant accuracy gains.

In the comparative analysis, several machine and deep learning-based models trained on original 4DCT images exhibited limited ability to detect pulmonary ventilation defects. Specifically, U-Net achieved a mean Dice of 0.51/0.51 in PET/SPECT cohort; U-Net++ attained 0.62/0.57; Res-UNet reached 0.59/0.67; and the RF classifier yielded 0.58/0.52. The suboptimal performance of these models can be attributed to their limited capacity to capture temporal dependencies inherent in respiratory motion. The standard U-Net architecture is primarily designed for spatial feature extraction, focusing on local spatial correlations through convolutional filters [68, 110]. In our implementation, the 4DCT was input as a multi-channel volume, where each respiratory phase was treated as a separate channel. Such approach treats temporal information as static, thereby neglecting dynamic inter-phase dependencies. U-Net++ extends U-Net by incorporating nested and dense skip connections to enhance spatial feature propagation and multi-scale aggregation [97], while Res-UNet introduces residual connections to facilitate signal propagation across network depths [95]. Despite these architectural improvements, both models similarly ignore critical temporal transitions and dynamics during the respiratory cycle [111]. The RF model is designed for tabular or statistical input, which lacks intrinsic mechanisms to learn sequential or temporal relationships [112].

To provide explainability, temporal saliency maps were generated from the proposed neural radiomic sequence model as in Fig. 11. These maps highlight key radiomic features and critical time points that contributed most significantly to the model's predictions. Notably, exhalation phases exhibited consistently higher saliency than inhalation phases. This observation aligns with physiological reasoning: because exhalation is the inverse of inhalation, the full dynamic range of motion is encoded in one direction, rendering it sufficient to characterize locoregional function. From the temporal saliency maps and associated feature importance histogram, three key features were identified as informative predictors of pulmonary ventilation impairment: Intensity-based 10th percentile, Intensity-based 90th percentile, and GLRLM-based Run-Length Non-Uniformity. As shown in Fig. 13, by explicitly modeling the temporal evolution of radiomic features along the respiratory cycle, our framework successfully preserves the cyclic nature of pulmonary motion. The temporal trajectories of these features provide insight into the physiological and structural mechanisms underlying ventilation impairment.

- The Intensity-based 10th percentile represents the lower-bound voxel intensity distribution within a region [101, 113], often corresponding to regions with low tissue density and high air content [114-116]. In healthy lung regions, this feature remained relatively constant across phases, indicating stable air content and consistent low-density areas during the breathing cycle. Conversely, in compromised regions, this feature increased during exhalation and subsequently declined during inhalation, with a symmetric shape reflecting the modeled respiratory waveform. This behavior suggests air trapping or impaired gas exchange in dysfunctional areas, leading to transient increases in tissue density during exhalation. The elevated absolute values of this feature in compromised regions, compared to healthy ones, potentially due to the impaired regions retain more residual air or fluid content.

- The Intensity-based 90th percentile captures the upper-bound of the intensity distribution [101, 113], typically reflecting dense parenchymal structures [114-116]. In healthy tissue, the 90th percentile exhibited a modest increase during exhalation and a corresponding decrease during inhalation, consistent with normal respiratory expansion and contraction. In contrast, compromised regions showed a more pronounced rise in intensity during exhalation, with a delayed return during inhalation. This exaggerated behavior likely reflects reduced compliance and abnormal mechanical deformation in diseased lung parenchyma. Notably, the consistent elevation of this feature in compromised tissue throughout the respiratory cycle highlights the persistently higher density characteristic of regions with ventilation abnormalities.
- The GLRLM-based Run-Length Non-Uniformity is a texture descriptor that quantifies the variability of contiguous voxel intensities [101, 117]. Healthy lung tissue maintained a stable pattern with relatively high non-uniformity across all phases, indicative of spatially variable but functionally normal lung textures. In contrast, compromised regions displayed a temporally dynamic pattern, with a notable decrease in non-uniformity during exhalation followed by recovery during inhalation. This trend suggests that during exhalation, dysfunctional regions exhibit reduced structural complexity or increased homogeneity, possibly due to collapse or restricted deformation. The lower absolute values of this feature in compromised regions support the interpretation that disease-related consolidation or fibrosis leads to more uniform textures in affected areas, which differs from the heterogeneous structure typical of ventilated lung tissue.

Collectively, our findings demonstrate that the proposed radiomic sequence model effectively captures meaningful temporal variations in both intensity- and texture-based features between healthy and functionally impaired lung regions. As 4DCT becomes increasingly integrated into clinical workflows for thoracic radiotherapy and pulmonary function analysis [33, 50], the developed methodology offers a non-invasive, image-based surrogate for functional lung assessment. It holds promises for delineating functionally critical sub volumes in functional lung avoidance radiotherapy and may also serve as a quantitative tool for monitoring regional responses in chronic respiratory diseases such as asthma, emphysema, and COPD.

Although the proposed framework operates at the voxel level—treating each lung voxel as an independent sample and thus leveraging millions of data points—the overall patient cohort remains modest, comprising 25 PET and 20 SPECT cases. To address this limitation, a five-fold cross-validation strategy was implemented to ensure robust performance evaluation across the entire dataset. Nonetheless, the restricted cohort size imposes inherent limitations on the generalizability of our findings. Accordingly, this study should be regarded as a methodological proof-of-concept, prioritizing algorithmic innovation over large-scale clinical validation. Future investigations involving larger and more diverse patient populations will be crucial to assessing the reproducibility, robustness, and clinical utility of the proposed approach. It is also important to acknowledge the heterogeneity between the SPECT and PET cohorts in terms of respiratory phase resolution and spatial image fidelity. While the model consistently identified salient radiomic features and maintained predictive stability across both modalities, these outcomes were derived from datasets acquired under differing imaging protocols. Therefore, future cross-modality validation using standardized datasets will be necessary to confirm the model's applicability in multi-institutional and real-world clinical environments.

## 6. Conclusion

In this study, we proposed a novel explainable neural radiomic sequence model that integrates spatiotemporal continuity for the voxel-wise identification of compromised pulmonary ventilation based on 4DCT imaging. By explicitly modeling the temporal evolution of radiomic features along the respiratory cycle, our framework preserves the cyclic nature of pulmonary motion and enables quantitative characterization of ventilation dynamics at the voxel level. Furthermore, the incorporation of temporal saliency analysis allows for the identification of critical radiomic patterns and respiratory phases that drive the model's predictions, thereby providing mechanistic insight into the underlying physiological defects. Given the increasing adoption of 4DCT in clinical workflows, this methodology is readily generalizable to other respiratory-related spatiotemporal imaging tasks.

## Declaration of competing interest

The authors declare that they have no known competing financial interests or personal relationships that could have appeared to influence the work reported in this paper.

## Acknowledgments

The authors would like to thank Dr. John Kipritidis, Dr. Paul Keall, Dr. Shankar SIVA, Dr. Tokihiro Yamamoto, and Dr. Joseph Reinhardt for sharing the VAMPIRE dataset.

## Data Availability

All imaging data used in this study is obtained from the publicly available VAMPIRE dataset [9]. The original data sets can be downloaded via request from the VAMPIRE website: https://image-x.sydney.edu.au/vampire-challenge/.